\documentclass[aps,prd,reprint,groupedaddress]{revtex4-2}
\usepackage{hyperref}
\usepackage{bm}
\usepackage{color}
\usepackage{amssymb}
\usepackage{amsmath}
\usepackage{dcolumn}
\usepackage{colortbl}
\usepackage{graphicx}
\usepackage{mathrsfs}
\usepackage{multirow}
\usepackage{longtable}
\usepackage[table,xcdraw]{xcolor}

\usepackage{url}
\usepackage{ulem}
\usepackage{xeCJK}
\usepackage{natbib}
\usepackage{diagbox}
\usepackage{rotating}
\usepackage{enumitem}
\usepackage{booktabs}
\usepackage{threeparttable}
\usepackage[thicklines]{cancel}

\begin{document}

\title{Limits on dark matter existence in neutron stars from recent astrophysical observations and mass correlation analysis}

\author{Jing Fu Hu}
\affiliation
{College of Physics and Electronic Information Engineering, Qinghai Normal University, Xining 810000, China}

\author{Hang Lu}
\affiliation
{College of Physics and Electronic Information Engineering, Qinghai Normal University, Xining 810000, China}
\affiliation
{MOE Frontiers Science Center for Rare Isotopes, Lanzhou University, Lanzhou 730000, China}
\affiliation
{School of Nuclear Science and Technology, Lanzhou University, Lanzhou 730000, China}

\author{Bao Yuan Sun\footnote{%
		Contact author (Email: sunby@lzu.edu.cn)}}
\affiliation
{MOE Frontiers Science Center for Rare Isotopes, Lanzhou University, Lanzhou 730000, China}
\affiliation
{School of Nuclear Science and Technology, Lanzhou University, Lanzhou 730000, China}


\begin{abstract}
   Dark matter admixed neutron stars (DANSs) serve as a specific astrophysical laboratory for probing the features of dark matter (DM) and have emerged as a promising candidate for interpreting recent astrophysical observations (e.g., by NICER and LIGO/Virgo). Accurately constraining the internal DM content of DANSs is therefore of critical importance. In this work, we construct the equations of state (EoS) for DANS matter by employing twelve nuclear matter (NM) models within the covariant density functional (CDF) theory and a self-interacting fermionic model for DM. Using these EoSs as input, we solve the two-fluid Tolman-Oppenheimer-Volkov (TOV) equations to systematically investigate the influence of DM on the global properties of neutron stars (NSs). By incorporating recent observational constraints on NS properties, the maximum DM mass fraction $f_\chi^{\mathrm{max}}$ in DANSs is determined for each NM EoS model. 
   Our analysis reveals a strong linear correlation (Pearson coefficient $r=0.98$) between $f_\chi^{\mathrm{max}}$ and the maximum mass of a pure NS, $M_{\rm{NS}}^{\mathrm{max}}$, described by $f_{\chi}^{\mathrm{max}} = 0.22 M_{\mathrm{NS}}^{\mathrm{max}} - 0.44$. Leveraging this correlation and the observed NS maximum mass distribution, $P(M_{\text{NS}}^{\max} \mid \text{EM})$, we derive the probability distribution function (PDF) for the maximum DM mass, $P(M_{\chi}^{\max} \mid \text{EM})$, in DANSs. We find that at the 68\% confidence level, $M_{\chi}^{\mathrm{max}}=0.150^{+0.070}_{-0.051}\ M_{\odot}$. This quantitative constraint on the DM mass provides a critical prior for interpreting potential observational signatures of DANSs, such as anomalous tidal deformabilities and distinctive gravitational-wave signals.
\end{abstract}


\maketitle

\section{Introduction}
Cosmological observations have revealed that the universe is composed of approximately 4.9\% ordinary matter, 26.8\% dark matter (DM), and 68.3\% dark energy  \cite{aghanim2021corrigendum}. As one of the dominant non-baryonic components in the universe, the existence of DM has been firmly established through various astrophysical observations, including galactic rotation curves, galaxy clusters, large-scale cosmological structures, and gravitational lensing \cite{bergstrom2000non, Kilbinger2015RPP086901,TULIN2018PR730,Mandelbaum2018ARAA56}. Nevertheless, the microscopic nature of DM remains elusive, its particle mass, interaction mechanisms, and coupling with ordinary matter remain among the central unsolved problems in modern cosmology and particle physics. To explain the evidence for DM provided by multi-scale astrophysical observations, various candidates have been proposed, such as weakly interacting massive particles (WIMPs) \cite{Goldman1989PhysRevD,JUNGMAN1996PR,arcadi2018TEPJC}, axions \cite{ROSENBERG2000PhyRep}, and sterile neutrinos \cite{Dodelson1994PRL}. Among these, the WIMP scenario is considered a possible candidate, as it naturally yields a relic abundance consistent with current cosmological observations and offers the possibility of detection in terrestrial experiments through weak interactions. Beyond theoretical predictions, a wide range of ground-based experiments is actively searching for DM signatures through different strategies \cite{Lewin1996AP, Mayet2016PhyRep,arcadi2018TEPJC,YaNG2019PhysRevLett,Catena2020PhysRevRes,Arg2021RevModPhys,Billard2022RepProgPhys,Cui2022PhysRevLett,Dai2022PhysRevLett}. These include direct detection, aimed at observing recoil signals resulting from collisions between DM particles and target nuclei or electrons, and indirect detection, which seeks secondary products of DM annihilation or decay.

Neutron stars, owing to their extremely high densities and intense gravitational fields, are regarded as ideal laboratories for exploring the properties of DM \cite{glendenning1997compact,haensel2007neutron,de2010neutron}. NSs can capture DM through gravitational accretion, giving rise to systems in which NM and DM coexist, commonly referred to as DANSs \cite{Kouvaris2010PhysRevD,leung2011dark,Güver2014JCAP,Raj2018PhysRevD, Bell2019JCAP,bell2021PRL,Fujiwara2022PRD,busoni2022MUPB}. Beyond direct accretion mechanisms, researchers have proposed several microscopic processes that may enable the formation and accumulation of DM within NSs. These include, for instance, the neutron dark decay processes that produce scalar DM particles \cite{Fornal2018PhysRevLett,Ellis2018PhysRevD}, as well as DM particles could be produced via bremsstrahlung in neutron-neutron scatterings \cite{Nelson2019JCAP}. Studies indicate that the presence of DM can remarkably alter the global structure and bulk properties of NSs, such as their mass-radius relations \cite{Ellis2018PhysRevD,Das2020MNRAS,Das2022PhysRevD,Rutherford2023PhysRevD}, moments of inertia \cite{Louren2022PhysRevD}, and tidal deformabilities \cite{Leung2022PhysRevD,Louren2022PhysRevD}, thereby providing an additional avenue for probing DM. If DM interacts with baryonic matter solely through gravity, its distribution within a DANS may take two characteristic forms depending on the particle mass, self-interaction strength, and relative abundance of the DM \cite{CIARCELLUTI2011PLB695,Xiang2014PhysRevC}. Light or strongly self-interacting DM tends to form an extended dark halo, while heavier DM is more likely to concentrate into a compact dark core. Both of these configurations significantly affect the tidal deformability of DANSs \cite{Xiang2014PhysRevC,Ivanytskyi2020PhysRevD,Miao2022APJ,Liu2024PhysRevD}. These distribution characteristics not only reflect the intrinsic nature of DM particles but are also closely linked to the EoS of NS matter. Consequently, studying DANSs offers profound insights into both the behavior of dense matter and the fundamental physics of DM.

Recent advances in multimessenger astrophysical observations and analytical techniques have delivered increasingly precise  NSs data, such as mass-radius measurements from NICER \cite{Fonseca_2021,Riley_2021,Salmi_2024, Vinciguerra_2024,Choudhury_2024} and tidal deformability constraints from LIGO/Virgo \cite{Abbott2017PhysRevLett, Abbott2020APJab960f}. DANSs have been invoked to account for these observational results \cite{Das2019PhysRevD,Das2021PhysRevD,Rafiei2022PhysRevD}. Through  comparisons between theoretical predictions and observational results (e.g., mass-radius relations, moments of inertia, tidal deformability and thermal evolution profiles), they have placed constraints on the properties of DM particles and their interaction strengths \cite{bramante2013PRD,Miao2022APJ,Rutherford2023PhysRevD,Thakur2024PhysRevD,Fibger2024JPG1361-6471,Cipriani2025PRD123005,Arvikar2025PRD023021}. These analyses provide valuable references for direct DM detection experiments. Concurrently, distinct observational signatures of DANSs have been investigated, such as an anomalous tidal deformability-radius relationship \cite{Sun2024PhysRevD,Liu2024PhysRevD}, gravitational waves with increased frequency and duration \cite{Rafiei2022PhysRevD,Hong2024PhysRevD}, and dark matter halo effects on pulse profiles \cite{Shakeri2024PhysRevD}. Such distinct  signatures offer concrete means to identify DANSs in observations. However, it is worth noting that in current research, the results are influenced by the choice of NM EoS models and DM models \cite{Das2021PhysRevD,Sun2024PhysRevD}. This occurs when using observational results to constrain DM properties and predict the observational signatures of DANSs. Moreover, even with fixed NM and DM models, the specific properties of DM particles and their abundance in DANSs are key parameters affecting the results.

Given the influence of the NM EoS models and the DM models on research outcomes, a less model-dependent approach is desirable. A possible way to implement this method lies in recognizing that the DM content is a key parameter for analyzing the global properties of DANSs, while astrophysical observations \cite{Abbott2017PhysRevLett,Bassa_2017,Fonseca_2021,Riley_2021,Salmi_2024,Vinciguerra_2024,Choudhury_2024} can constrain these global properties. It would therefore be motivated to determine whether a correlation exists between the DM content and these observational constraints, this could reduce uncertainties stemming from the choice of NM EoS models and DM models, thereby  providing a method for constraining the DM content in DANSs via observations. Motivated by this prospect, this work aims to explore correlations between the DM content in DANSs and the global properties (e.g., maximum mass) of NSs. To reduce model‑dependent effects, different NM EoS models are adopted to compute the global properties of NSs, and the results are compared with recent observations. To this end, we employ twelve NM EoS models within the Covariant Density Functional (CDF) theory, which is widely used in NSs studies because it provides an accurate description of finite nuclei and can be extended to construct EoS for dense matter \cite{ReinhardRPP1989,RING1996193,Bender2003RevModPhys,VRETENAR2005101,MENG2006470,Meng2016WS}.  The dark matter EoS is constructed using a model of non-annihilating, self-interacting fermions \cite{Narain2006PhysRevD}.  By numerically solving the two-fluid Tolman–Oppenheimer–Volkoff (TOV) equations \cite{Tolman1939PhysRev,Oppenheimer1939PhysRev,kodama1972PTP,comer1999PRD,sandin2009AP,CIARCELLUTI2011PLB},  the global properties of pure NSs and DANSs are first analyzed and then compared with observational results. Following this, a possible correlation between the DM content in DANSs and the global properties of pure NSs and DANSs is investigated. The proposed approach is also intended to allow for the subsequent inclusion of other DM models in future studies.

The paper is organized as follows. In Sec.~\ref{sec:framework} details the theoretical framework for NM and DM, including numerical implementation of two-fluid TOV equations. In Sec.~\ref{sec:analysis} analyze the effects of $f_{\chi}$ on the global properties of NSs and discuss the upper limit on the DM content imposed by observations. Finally, a summary will be given in Sec.~\ref{sec:conclusions}. We use units in
which $\hbar = c = G = 1$.

\section{Theoretical Framework}\label{sec:framework}
\subsection{Equation of state for nuclear matter}
Research on DM is fundamentally dependent on a reasonable description of NM properties. To establish this foundation, we begin with a concise introduction to the theoretical framework of pure NM EoS models before proceeding to a discussion of DM. The CDF theory has been extensively applied in nuclear structure and astrophysics studies, owing to its remarkable success in describing both single-particle and collective properties of finite nuclei \cite{ReinhardRPP1989,RING1996193,Bender2003RevModPhys,VRETENAR2005101,MENG2006470,Meng2016WS}. In this work, based on the meson-exchange diagram of the nuclear force, we conduct a theoretical investigation using three representative CDF models. Specifically, we employ the nonlinear relativistic mean field (NLRMF) \cite{BOGUTA1977NPA,Lalazissis1997PhysRevC}, density-dependent relativistic mean field (DDRMF) \cite{Brockmann1992PRL}, and density-dependent relativistic Hartree-Fock (DDRHF) \cite{LONG2006PLB,LONG2007PhysRevC} approaches, all derived from the following Lagrangian density
\begin{align}\label{Lagrangian}
	\mathcal{L} &= \mathcal{L}_{B} + \mathcal{L}_{m} + \mathcal{L}_{\rm{int}} + \mathcal{L}_{\rm{NL}},
\end{align}
the terms $\mathcal{L}_{B}$ and $\mathcal{L}_{m}$ represent the free Lagrangians of baryons, leptons, and mesons, respectively, while $\mathcal{L}_{\rm{int}}$ describes the interactions among baryon fields mediated by meson exchange \cite{CHIN1974PLB}. In nonlinear RMF models, an additional term, $\mathcal{L}_{\rm{NL}}$ is introduced to incorporate meson self-interactions as well as nonlinear mixing effects \cite{BOGUTA1977NPA}. Starting from the Lagrangian density $\mathcal{L}$ given in Eq. \ref{Lagrangian}, the corresponding effective Hamiltonian operator $\hat{H}$ is derived through a generalized Legendre transformation. The system energy functional is subsequently obtained by evaluating the expectation value of the Hamiltonian operator with respect to the ground state $|\Phi_{0}\rangle$, following the standard procedure \cite{CHIN1974PLB,BOGUTA1977NPA}.

In $\beta$-stable NSs matter consisting of nucleons $(n,p)$, electrons $e$ and muons $\mu$, the condition of chemical equilibrium dictates that
\begin{align}
	\mu_{p}=\mu_{n}-\mu_{e}, \quad \mu_{\mu}=\mu_{e},
\end{align}
the chemical potentials of neutrons, protons, muons and electrons, $\mu_{n}$, $\mu_{p}$, $\mu_{\mu}$, and $\mu_{e}$ are determined from the relativistic energy–momentum relation evaluated at the Fermi momentum $k_{F}$ \cite{Sun2008PhysRevC}. In addition, the system is constrained by baryon density conservation and the condition of charge neutrality, namely
\begin{align}
	\rho_{b}=\rho_{n}+\rho_{p}, \quad \rho_{p}=\rho_{\mu}+\rho_{e}.
\end{align}
Based on the above constraints, the energy density of NSs matter can be further determined
\begin{align}
	\varepsilon_{\rm{NS}}=\sum_{i=n,p,e,\mu} \varepsilon_{k,i}+\sum_{\phi=\sigma,\omega,\rho,\pi}\left(\varepsilon_{\phi}^{D}+\varepsilon_{\phi}^{E}\right),
\end{align}
in the energy functional, $\varepsilon_{k}$ denotes the kinetic energy density, while $\varepsilon_{\phi}^{D}$ and $\varepsilon_{\phi}^{E}$ represent the direct (Hartree) and exchange (Fock) contributions to the potential energy density, respectively. Specifically, the nucleonic component of the kinetic energy density and the potential energy density can be written in the following form
\begin{align}
	\varepsilon_{k,N} &= \sum_{i=n,p} \frac{1}{\pi^2} \int_0^{k_{F,i}} p^2 dp (p\hat{P} + M \hat{M}),\\
	\varepsilon_{\sigma}^{D}&=-\frac{1}{2}\frac{g_{\sigma}^{2}}{m_{\sigma}^{2}}\rho_{s}^{2}, ~ \varepsilon_{\omega}^{D}=\frac{1}{2}\frac{g_{\omega}^{2}}{m_{\omega}^{2}}\rho_{b}^{2}, ~ \varepsilon_{\rho}^{D}=\frac{1}{2}\frac{g_{\rho}^{2}}{m_{\rho}^{2}}\rho_{b3}^{2},\\
	\varepsilon_{\phi}^{E} &= \frac{1}{2}\frac{1}{(2\pi)^{4}}\sum_{\tau,\tau^{\prime}}\delta_{\tau\tau^{\prime}}\int pp^{\prime} dpdp^{\prime}\left[A_{\phi}\left(p, p^{\prime}\right)\right.\notag \\
&\quad \left.+\hat{M}(p)\hat{M}\left(p^{\prime}\right) B_{\phi}\left(p, p^{\prime}\right)\right. \notag \\
&\quad \left.+\hat{P}(p)\hat{P}\left(p^{\prime}\right) C_{\phi}\left(p, p^{\prime}\right)\right],
\end{align}
here, $\hat{P}$, and $\hat{M}$ denote the hatted quantities, $p$ is momentum, while $\rho_{s}$, $\rho_{b}$, and $\rho_{b3}$ represent the scalar density, the baryon density, and its third component, respectively. The term $\delta_{\tau\tau^{\prime}}$ stands for the isospin-related factor, and $A_{\phi}$, $B_{\phi}$, and $C_{\phi}$ are angular integral coefficients.  The explicit expressions for these quantities can be found in Ref.~\cite{Sun2008PhysRevC}. It is worth noting that, in the NLRMF model, the additional contribution $\varepsilon_{\rm{NL}}$ originating from meson nonlinear self-couplings is absorbed into the Hartree term $\varepsilon_{\phi}^{D}$ of the potential energy
\begin{align} 
	\varepsilon_{\rm{NL}}=\frac{1}{3}g_{2}\sigma^{3}+\frac{1}{4}g_{3}\sigma^{4}-\frac{1}{4}c_{3}\omega^{4}-\frac{1}{2}\Lambda_{v}g_{\omega}^{2}g_{\rho}^{2}\omega^{3}\rho^{2}.
\end{align}
Furthermore, in this work, leptons are treated as a free Fermi gas, under the assumption that they do not interact with nucleons or mesons. Thus, the kinetic energy of the leptons can be written as
\begin{align}
	\varepsilon_{k,\lambda} &= \frac{1}{\pi^2} \int_0^{k_{F,\lambda}} p^{2} dp \sqrt{p^{2}+m_{\lambda}^{2}}.
\end{align}
The pressure of the NSs system can be obtained from the thermodynamic relation
\begin{align}
	P_{NS}\left(\rho_{b}\right)=\rho_{b}^{2} \frac{d}{d\rho_{b}} \left(\frac{\varepsilon_{\rm{NS}}}{\rho_{b}}\right)=\sum_{i=n,p,e,\mu}\rho_{i}\mu_{i}-\varepsilon_{\rm{NS}}.
\end{align}
The crust structure in the low-density region ($\rho_{b} < 0.08~\rm{fm}^{-3}$) is described using the Baym-Pethick-Sutherland (BPS) EoS \cite{Baym1971ApJ} for the outer crust and the Baym-Bethe-Pethick (BBP) EoS \cite{Gordon1971NuPhA} for the inner crust.

\subsection{Equation of State for dark matter}
 In this work, the DM candidate is a weakly interacting massive particle (WIMP), modeled as a non-annihilating, self-interacting fermion. WIMPs are selected primarily because their thermal freeze-out mechanism accounts for the observed DM relic abundance \cite{kolb2018early,Steigman2012PhysRevD,arcadi2018TEPJC}. Their weak-scale interactions enable detection in direct and indirect experiments \cite{Lewin1996AP,Mayet2016PhyRep}, while also allowing them to be captured and accumulated in dense astrophysical environments like NSs \cite{Güver2014JCAP}. We assume DM particle mass to be between MeV and GeV the specific formulation is taken from Ref.~\cite{Narain2006PhysRevD}. Here, we present its equation of state, i.e., energy density and pressure

\begin{align}
    &\varepsilon_{\chi} = \frac{1}{\pi^{2}} \int_{0}^{k_{F}} p^{2} dp\sqrt{p^{2} + m_{\chi}^{2} } \,  +  \frac{\rho_{\chi}^{2}}{m_{I}^{2}} , \\
    &P_{\chi} = \frac{1}{3\pi^{2}} \int_{0}^{k_{F}}dp \frac{p^{4}}{\sqrt{p^{2} + m_{\chi}^{2}}} \,  +  \frac{\rho_{\chi}^{2}}{m_{I}^{2}},
\end{align}
where $m_{\chi}$ is the mass of DM particles, and the Fermi momentum $k_{F}$ is related to the number density $\rho_{\chi}$ by $k_{F} = (3\pi^{2} \rho_{\chi})^{1/3}$, $m_I$ is the interaction mass scale. By
denoting $x \equiv k_{F}/m_{\chi}$ and $y \equiv m_{\chi}/m_{I}$, the energy density and pressure are determined as
\begin{align}
\varepsilon_{\chi} &= \frac{m_{\chi}^{4}}{8\pi^{2}}\left[(2x^{3}+x)\sqrt{1+x^{2}}-\sinh^{-1}(x)\right] \notag \\
&\quad +\frac{m_{\chi}^{4}y^{2}x^{6}}{(3\pi^{2})^{2}}, \\
P_{\chi} &= \frac{m_{\chi}^{4}}{24\pi^{2}}\left[(2x^{3}-3x)\sqrt{1+x^{2}}+3\sinh^{-1}(x)\right] \notag \\
&\quad +\frac{m_{\chi}^{4}y^{2}x^{6}}{(3\pi^{2})^{2}}.
\end{align}
For both weak interaction DM (with a typical scale of $m_I \sim 300\ \mathrm{GeV}$, comparable to the $W$ or $Z$ boson masses) and strongly interacting DM (where $m_I \sim 100\ \mathrm{MeV}$ is assumed based on chiral perturbation theory), the interaction mass scales are adopted from Ref.~\cite{Narain2006PhysRevD}. For this work, we fix the DM particle mass at $m_\chi = 1000$ MeV and the dimensionless self-interaction strength at $y=0.1$. These values lie within the mass and weak-interaction scale of WIMPs \cite{leung2011dark,Narain2006PhysRevD}.

\subsection{Two-fluid TOV Equations}

The two-fluid TOV equations are employed to determine the structural configuration of a DANS
\cite{Tolman1939PhysRev,Oppenheimer1939PhysRev,kodama1972PTP,comer1999PRD,sandin2009AP,CIARCELLUTI2011PLB}, in which NM and DM interact exclusively via gravity, while each component independently conserves its own energy-momentum. The total energy density $\varepsilon(r)$ and pressure $P(r)$ are given by the linear superposition of contributions from both components, with the subscript indices N and $\chi$ standing for the NM and DM components, respectively

\begin{align}  
	P(r)=P_{\rm{N}}(r)+P_{\chi}(r) \label{eq:pressure}, \\
	\varepsilon(r)=\varepsilon_{\rm{N}}(r)+\varepsilon_{\chi}(r). \label{eq:energy_density}
\end{align}
The coupled TOV equations governing the system are formulated as follows
\begin{align}   
	\frac{dP_{\rm{N}}}{dr} &= -(P_{\rm{N}} + \varepsilon_{\rm{N}}) \frac{4\pi r^{3}(P_{\rm{N}} + P_{\chi}) + M(r)}{r(r - 2M(r))} \label{eq:dpb}, \\
	\frac{dP_{\chi}}{dr} &= -(P_{\chi} + \varepsilon_{\chi}) \frac{4\pi r^{3}(P_{\rm{N}} + P_{\chi}) + M(r)}{r(r - 2M(r))} \label{eq:dpx},
\end{align}

\begin{align} 
	\frac{dM_{\rm{N}}}{dr} &=4\pi \varepsilon_{\rm{N}} r^{2} \label{eq:mass_derivative}, \\ 
	\frac{dM_{\chi}}{dr} &=4\pi \varepsilon_{\chi} r^{2} \label{eq:mass_divative}. 
\end{align}
The total gravitational mass is defined as the sum of the two components evaluated at their respective radial boundaries 
\begin{align}
	M_{\rm{T}}=M_{\rm{N}}(R_{\rm{N}})+M_{\chi}(R_{\chi}).
\end{align}
The EoS for nuclear matter and dark matter are used as input parameters, where $P_{N,c}$ and $P_{\chi,c}$ represent the central pressures of the NM and DM components, respectively, as initial conditions. Numerically solve the two-fluid TOV equations until the pressure on the surface of each matter is zero, i.e., $P_{\rm{N}}(r) = 0$ and $P_{\chi}(r) = 0$. This gives the radii of the nuclear matter
component $R_{\rm{N}}$ and the DM component $R_{\chi}$. The gravitational
masses of the two components are then $M_{\rm{N}}(R_{\rm{N}})$ and $M_{\chi}(R_{\chi})$,
respectively. Due to the nonluminous nature of DM, the observable star radius $R_{T}$ is identified with the NM radius $R_{\rm{N}}$. Hereafter, we denote $f_{\chi} = M_{\chi}(R_{\chi}) / M_{\rm{T}}$ as the DM mass fraction. For pure NSs, the gravitational mass and its corresponding radius are denoted as $M_{\mathrm{NS}}$ and $R_{\mathrm{NS}}$, respectively.

\section{Results and Discussion}\label{sec:analysis}
\subsection{Effects of Dark Matter on Neutron Star Properties}
\begin{figure}[htbp]
	\centering
	\includegraphics[width=0.48\textwidth]{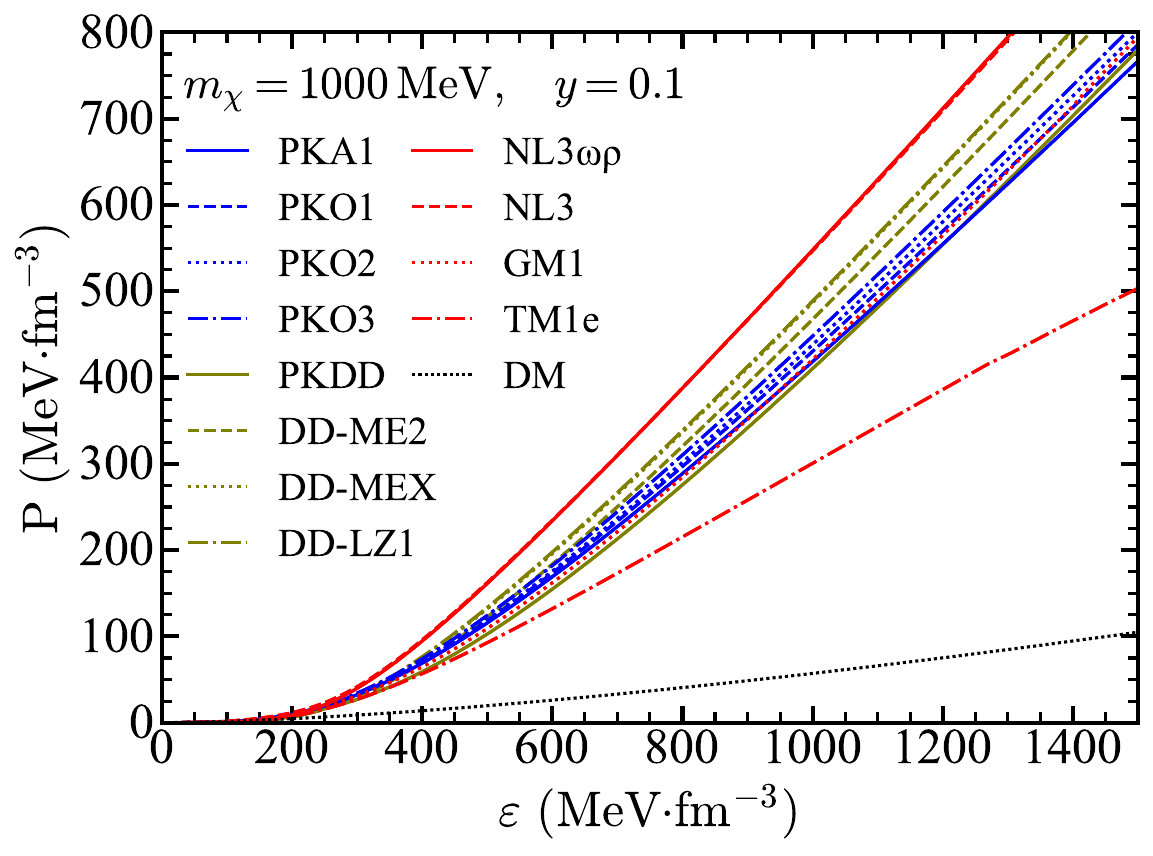}
	\caption{The pressure as a function of the energy density for both $\beta$-equilibrated nuclear matter ($P_{\rm{N}}$ vs $\varepsilon_{\rm{N}}$) and the fermionic dark matter ($P_{\chi}$ vs $\varepsilon_{\chi}$). The CDF models are selected from DDRHF models (blue), DDRMF models (olive green), and NLRMF models (red). The dark matter model (labeled as DM) takes the value of $m_{\chi}=1000$ MeV, and the dimensionless self-interaction strength $y=0.1$.}\label{fig:1}
\end{figure}

In this work, the global properties of DANSs are investigated using the two‑fluid TOV equation, with NM EoS models constructed from twelve CDF effective Lagrangians. These include the DDRHF models PKA1 \cite{LONG2007PhysRevC}, PKO1 \cite{LONG2006PLB}, PKO2, and PKO3 \cite{Long2008EpL12001}, the DDRMF models PKDD \cite{long2004new}, DD-ME2 \cite{lalazissis2005new},  DD-MEX \cite{taninah2020parametric} and DD-LZ1 \cite{wei2020novel}, as well as the NLRMF models NL3$\omega\rho$ \cite{Horowitz2001}, NL3 \cite{Lalazissis1997PhysRevC}, GM1 \cite{Glendenning1991PhysRevLett} and TM1e \cite{Shen2020TAAS}. In our previous work, these models have been applied in related studies to investigate NS properties and NM behavior
\cite{SUN2010PLB,Long2012PhysRevC,Qi2016RAA,Liu2018PhysRevC,qian2018SCP,Yang2019PhysRevC,Yang2021PhysRevC,Ding2025PRD}. The EoSs for $\beta$-equilibrated NM based on these CDF effective Lagrangians are presented in Fig.~\ref{fig:1}, where the blue, olive green, and red curves correspond to the DDRHF, DDRMF, and NLRMF models respectively.
At high densities, the stiffness of the EoS varies across the models, a variation associated with NS properties such as incompressibility and symmetry energy. Among the set considered, the stiffest EoS is provided by the NL3 model. Also included in Fig.~\ref{fig:1} is the EoS for a self‑interacting fermionic DM model, with a particle mass of $m_{\chi}=1000$ MeV and a dimensionless self‑interaction strength of $y=0.1$. The DM EoS is substantially softer than all the NM EoSs, implying that an admixture of such DM could alter the global properties of NSs.

\begin{figure}[htbp]
	\centering
	\includegraphics[width=0.48\textwidth]{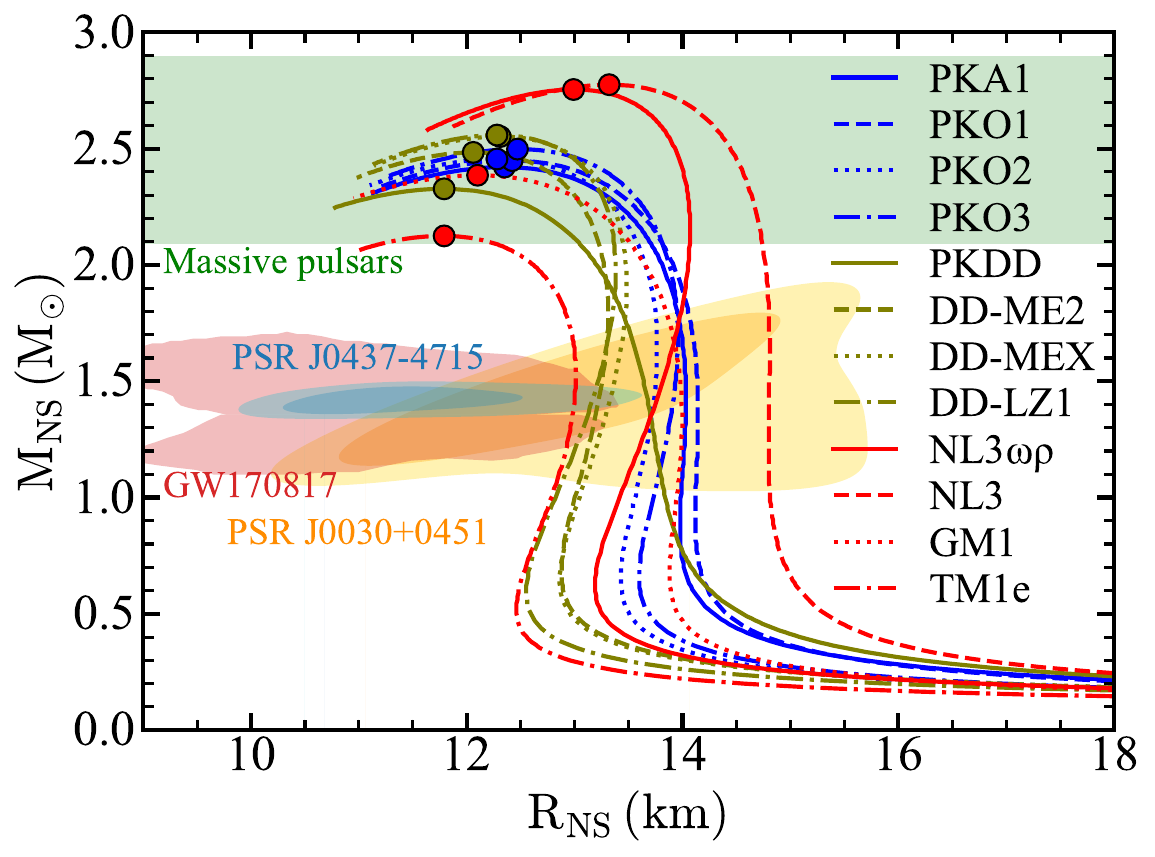}
	\caption{Mass-radius relations of pure neutron stars obtained by the selected CDF equations of state. The shaded regions indicate several astrophysical observational constraints, including the distribution of the pulsar mass limit (green) from electromagnetic observations \cite{romani2022psr}, the mass-radius measurements from the NICER mission for PSR J0030+0451 (yellow) \cite{Vinciguerra_2024} and the brightest known millisecond pulsar PSR J0437-4715 (blue) \cite{Choudhury_2024}, the gravitational wave event GW170817 (red) \cite{Abbott2017PhysRevLett}.}\label{fig:2}
\end{figure}

To investigate the influence of DM on NSs, the mass–radius relations of pure NSs are calculated using the twelve NM EoS models presented in Fig.~\ref{fig:1}. These relations, together with recent observational constraints on NSs mass and radius, are presented in Fig.~\ref{fig:2}. Analysis of Fig.~\ref{fig:1} and Fig.~\ref{fig:2} reveals a consistent trend across the considered models that stiffer NM EoS models generally predicts a larger maximum mass for NSs, $M_{\mathrm{NS}}^{\mathrm{max}}$. For example, the highest value of $M_{\mathrm{NS}}^{\mathrm{max}} = 2.775~M_{\odot}$ is yielded by the NL3 model, as marked by a solid dot in Fig.~\ref{fig:2}, while the maximum masses predicted by the other models are listed in Table~\ref{table:1}. 
The green shaded region in Fig.~\ref{fig:2} represents the probability distribution function (PDF) $P(M_{\text{NS}}^{\max} \mid \text{EM})$ \cite{romani2022psr} for the  mass of supermassive pulsars (e.g., PSR J0740+6620\cite{Fonseca_2021}, and PSR J0952-0607\cite{Bassa_2017}) derived from electromagnetic (EM) observations, and in our investigation we refer to this as the \textit{“MaxM”} constraints. In this work, it is required that the maximum mass predicted by each employed NM EoS model satisfies this \textit{MaxM} constraints. Simultaneously, the mass–radius constraints provided by PSR~J0030+0451 \cite{Vinciguerra_2024}, the GW170817 binary neutron star merger event \cite{Abbott2017PhysRevLett}, and PSR~J0437-4715 \cite{Choudhury_2024} are also presented in Fig.~\ref{fig:2}, which we collectively refer to as the \textit{“TycM”} constraints in this investigation. Some NM EoS models predictions fail to meet the observational radius constraints provided by GW170817 and PSR~J0437-4715.
Given that the DM EoS shown in Fig.~\ref{fig:1} is softer than those of the NM EoS models, we further consider the admixture of DM into these models. This enables the analysis of DM influence on NS global properties, and compare the resulting outcomes with \textit{TycM} and \textit{MaxM} constraints.

\begin{figure}[htbp]
	\centering
	\includegraphics[width=0.48\textwidth]{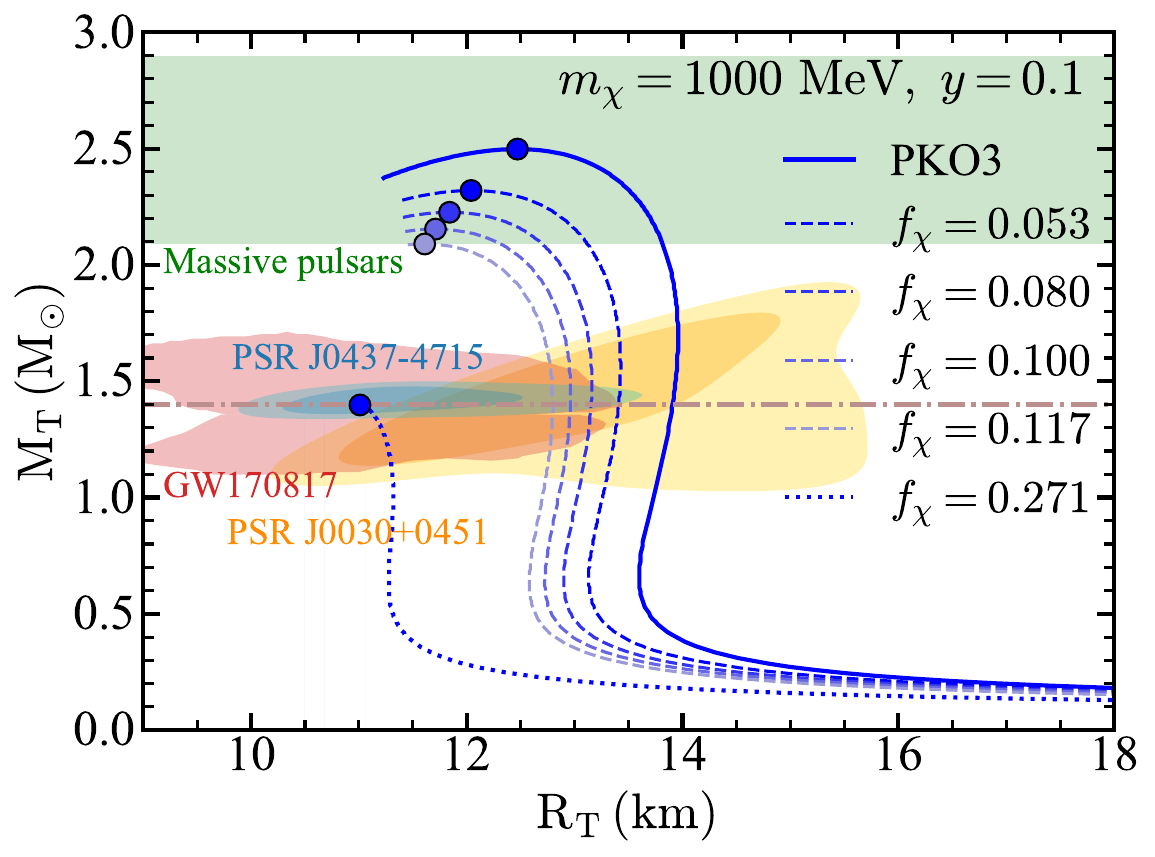}
	\caption{Mass-radius relations (the total mass $M_{\rm{T}}$ vs the observable radius $R_{\rm{T}}$) of dark matter admixed neutron stars with different dark matter fraction $f_{\chi}$ (dashed lines with graduated colors), using the CDF model PKO3. For comparison, the case of pure neutron stars is given by the solid line. The dash-dotted line marks the fixed mass of $M_{\rm{T}}=1.4~M_{\odot}$.}\label{fig:3}
\end{figure}

An analysis of the mass-radius relations for DANSs is conducted through calculations with varying DM mass fractions, $f_{\chi}$. A fermionic DM model is adopted, with a DM particle mass $m_{\chi} = 1000$ MeV and a dimensionless self‑interaction strength $y = 0.1$. Fig.~\ref{fig:3} shows that the predicted maximum mass and corresponding radii of DANSs are lower than those of pure NSs and decrease as $f_{\chi}$ increases, this result on the impact of such DM on the global properties of NSs are consistent with the conclusions reported in Refs.~\cite{Miao2022APJ,Liu2024PhysRevD}.  For illustration, the mass-radius relations for the PKO3 model are presented in Fig.~\ref{fig:3}. When \textit{MaxM} constraints are imposed, we find that there exists an upper limit on the DM mass fraction $f_{\chi}$ in the DANSs, and for the PKO3 model, this limit is $f^{\rm max}_{\chi}=0.117$.  If the DM mass fraction exceeds this limit, the resulting mass-radius relation can no longer satisfy \textit{MaxM} constraints. The values of $f^{\rm max}_{\chi}$
and the corresponding maximum total mass $M^{\rm max}_{\rm{T}}$ 
of the DANS for each model, obtained under this \textit{MaxM} constraints, are compiled in Table~\ref{table:1}. The results show that the upper limit on the DM mass fraction $f^{\rm max}_{\chi}$ constrained by \textit{MaxM}, depends on the specific NM EoS models employed.

\begin{figure}[htbp]
	\centering
	\includegraphics[width=0.48\textwidth]{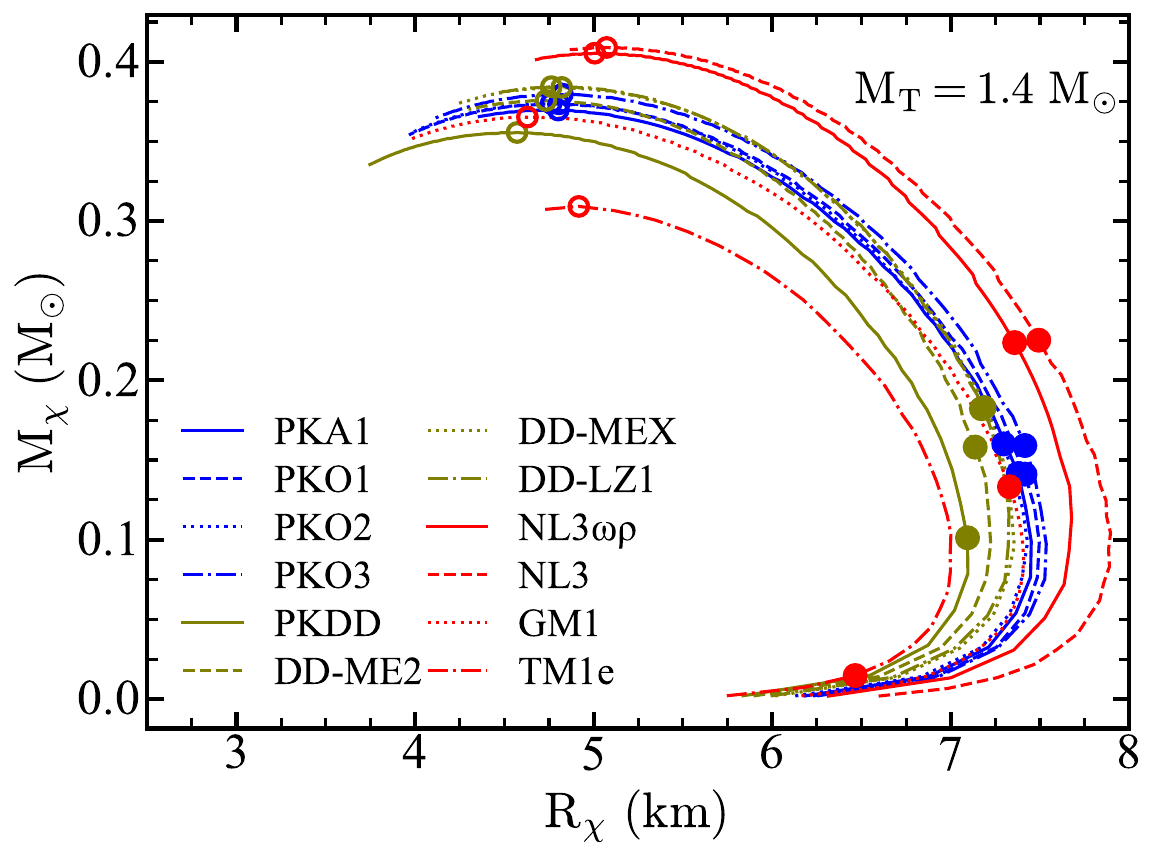}
	\caption{For a DANSs with fixed total mass of $M_{\rm{T}}=1.4\,M_\odot$, the mass of the dark matter component $M_\chi$ as a function of DM radius $R_\chi$, which is evolved with increasing DM mass fraction $f_{\chi}$, the results are given by different CDF models. The maximum DM masses which are permissible from  \textit{TycM} constraints $M_{\mathrm{\chi}}^{\mathrm{*1.4}}$ (open circles) and  constrained by the \textit{MaxM} $M_{\mathrm{\chi}}^{1.4}$ (solid circles) are marked.}\label{fig:4}
\end{figure}

The upper limits on the DM content within DANSs, as established by \textit{MaxM}  constraints, are presented in Fig.~\ref{fig:3} and Table~\ref{table:1}. We now present an investigation of the upper limit on the internal DM content under the  \textit{TycM} constraints for a DANS with a typical NS mass of $1.4~M_{\odot}$. To this end, we investigate the evolution of the mass-radius relation of the DM component by fixing the total mass of the DANS,  $M_{\rm{T}}$, at the typical NS mass of $1.4~M_{\odot}$, and progressively increasing $f_{\chi}$, thereby tracing the correlated evolution of $M_{\chi}$ and $R_{\chi}$, as shown in Fig.~\ref{fig:4}. The  \textit{TycM} constraints upper limits $M_{\chi}^{*1.4}$ are indicated by hollow dots, while solid dots denote the DM masses $M_{\chi}^{1.4}$ derived under \textit{MaxM} constraints. The predicted values of $M_{\chi}^{*1.4}$ and $M_{\chi}^{1.4}$ for different NM EoS models are compiled in Table~\ref{table:1}. 
In Fig.~\ref{fig:3}, the DM masses under \textit{MaxM} and  \textit{TycM} constraints correspond to the intersections of the mass‑radius curves for $f_{\chi}^{\rm{max}}=0.271$ and 
$f_{\chi}^{\rm{max}}=0.117$ with the $1.4~M_{\odot}$ contour, respectively. The results show that the  \textit{TycM} constraints upper limit of the DM mass is dependent on the choice of the NM EoS models. Given that the determined upper limits for the DM content under both the \textit{MaxM} and the  \textit{TycM} constraints are influenced by the choice of NM EoS models, it is necessary to search for possible correlations to reduce this model dependence.

In this work, when the DM particle mass is set to 
$m_{\chi}=1000$ MeV and the self-interaction strength to 
$y=0.1$, the DM exists in the form of a dark core within the DANSs. This is demonstrated by the data in Table~\ref{table:1}, which show that across all models, the DM radius $R^{\max}_{\chi}$ is substantially less than the total radius $R^{\rm{max}}_{\rm{T}}$. For example, in the PKO3 model, the DM core radius is only $R_{\chi}^{\rm{max}}= 4.320$ km compared to the total $R_{\rm{T}}^{\max} = 11.610$ km.

\begin{table*}[htbp]\caption{\label{tab:eos-comparison}
The maximum mass $M_{\rm{NS}}^{\rm{max}}$ ($M_{\odot}$) and the corresponding radius $R_{\rm{NS}}^{\rm{max}}$ (km) for a pure neutron stars calculated by different CDF models. To their right, the maximum value of the dark matter fraction $f_{\chi}^{\rm{max}}$ is given for dark matter admixed neutron stars, which is constrained by astrophysical observations \cite{Abbott2017PhysRevLett,romani2022psr,Choudhury_2024,Vinciguerra_2024}. Correspondingly, the predict results of the total mass $M_{\rm{T}}^{\rm{max}}$, the observable radius $R_{\rm{T}}^{\rm{max}}$, the DM mass $M_{\chi}^{\rm{max}}$ and DM radius $R_{\chi}^{\rm{max}}$ are obtained for the DANSs case of maximum mass limit. For the case of 1.4~$M_{\odot}$ DANSs, the predicted maximum DM masses $M_{\chi}^{1.4}$ under \textit{MaxM} constraints are then compared with those permissible from  \textit{TycM} constraint $M_{\rm{\chi}}^{*1.4}$, see Fig.~\ref{fig:4} for details.}
\begin{ruledtabular} 
\renewcommand{\arraystretch}{1.5}
\doublerulesep 0.1pt \tabcolsep 1.5pt
\begin{tabular}{@{}lccccccccc@{}}
\multicolumn{1}{c}{} & \multicolumn{2}{c}{NS} & \multicolumn{7}{c}{DANS} \\
\cline{2-3} \cline{4-10}  
EoS &
$M_{\rm{NS}}^{\rm{max}}$ &
$R_{\rm{NS}}^{\rm{max}}$ &
$f_{\chi}^{\rm{max}}$ &
$M_{T}^{\rm{max}}$ &
$R_{T}^{\rm{max}}$ &
$M_{\rm{\chi}}^{\rm{max}}$ &
$R_{\rm{\chi}}^{\rm{max}}$ &
$M_{\rm{\chi}}^{1.4}$ &
$M_{\rm{\chi}}^{*1.4}$ \\
\hline
PKA1 & 2.421 & 12.340 & 0.099 & 2.090 & 11.610 & 0.207 & 4.460 & 0.138 & 0.370 \\
PKO1 & 2.449 & 12.400 & 0.105 & 2.090 & 11.630 & 0.220 & 4.410 & 0.147 & 0.374 \\
PKO2 & 2.456 & 12.280 & 0.108 & 2.090 & 11.480 & 0.226 & 4.360 & 0.151 & 0.374 \\
PKO3 & 2.498 & 12.460 & 0.117 & 2.090 & 11.610 & 0.245 & 4.320 & 0.164 & 0.380 \\
PKDD & 2.327 & 11.780 & 0.076 & 2.090 & 11.240 & 0.159 & 4.370 & 0.106 & 0.356 \\
DD-LZ1 & 2.558 & 12.280 & 0.130 & 2.090 & 11.350 & 0.272 & 4.120 & 0.182 & 0.384 \\
DD-ME2 & 2.484 & 12.060 & 0.116 & 2.090 & 11.230 & 0.242 & 4.210 & 0.163 & 0.376 \\
DD-MEX & 2.552 & 12.300 & 0.129 & 2.090 & 11.390 & 0.270 & 4.120 & 0.181 & 0.384 \\
NL3 & 2.775 & 12.470 & 0.161 & 2.090 & 12.470 & 0.336 & 4.310 & 0.224 & 0.409 \\
GM1 & 2.386 & 12.100 & 0.092 & 2.090 & 11.430 & 0.192 & 4.390 & 0.128 & 0.365 \\
TM1e & 2.125 & 11.790 & 0.011 & 2.090 & 11.690 & 0.023 & 4.240 & 0.015 & 0.309 \\
NL3$\omega\rho$ & 2.754 & 13.000 & 0.158 & 2.090 & 12.050 & 0.330 & 4.250 & 0.222 & 0.405 \\
\end{tabular}
\end{ruledtabular}
\label{table:1}
\end{table*}

\subsection{Limit on the dark matter component in DANSs}
\begin{figure}[htbp]
	\centering
	\includegraphics[width=0.48\textwidth]{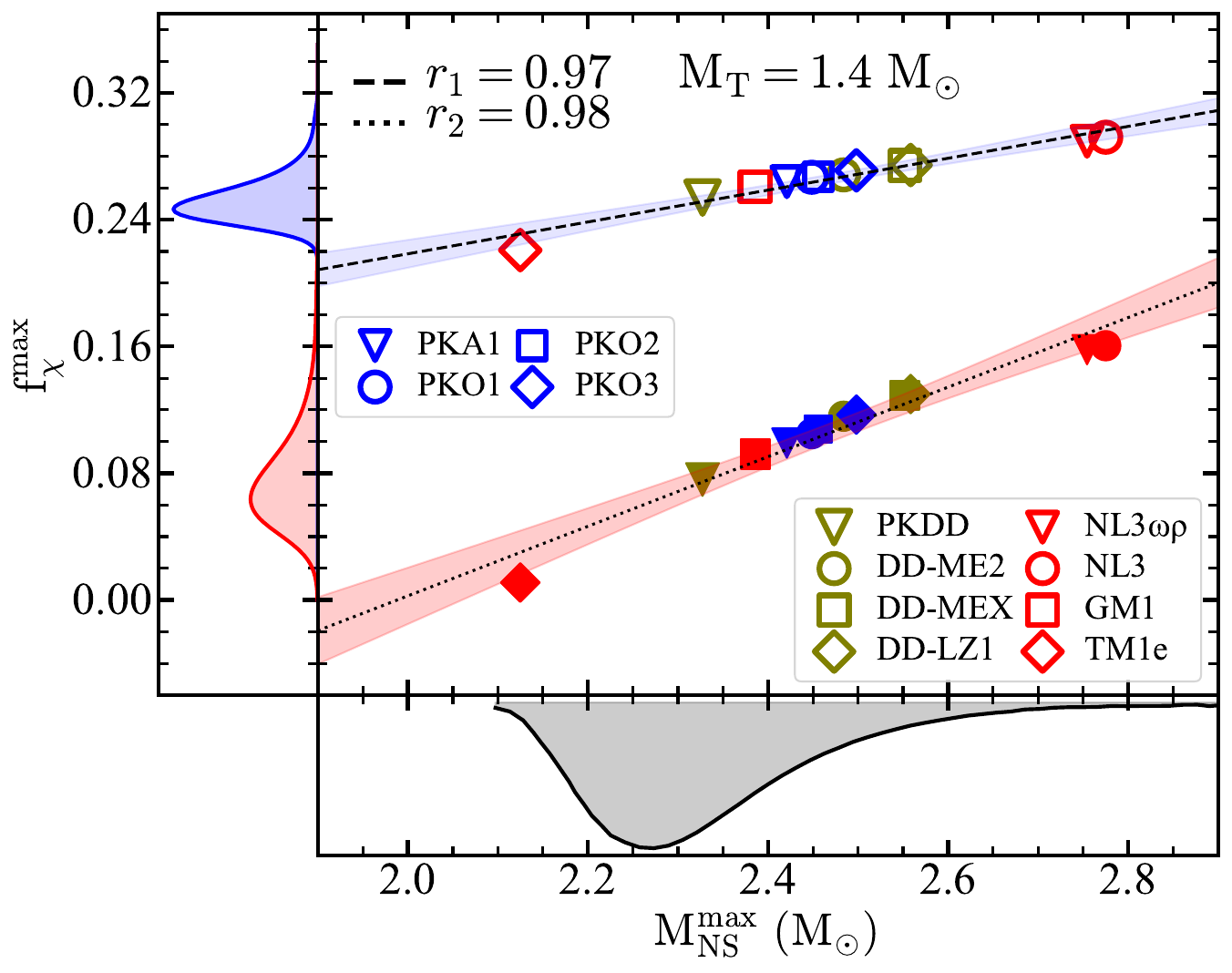}
	\caption{Linear correlation between the maximum dark matter mass fraction (derived from $1.4\,M_\odot$ DANSs) and the maximum mass of a pure neutron stars. The dashed line represents the  \textit{TycM} constraints limit $f_{\chi}^{\mathrm{*max}}$, with a Pearson correlation coefficient of $r_1 = \mathrm{0.97}$. The blue shaded area denotes the confidence interval. The dotted line represents the \textit{MaxM} constraints limit $f
	_{\chi}^{\mathrm{max}}$, with a Pearson correlation coefficient of $r_1 = \mathrm{0.98}$. The red shaded area denotes the confidence interval.
	Bottom panel: the probability distribution function of maximum mass limit $P(M_{\text{NS}}^{\max} \mid \text{EM})$ for a pure neutron star constrained by \textit{MaxM} \cite{romani2022psr}.
	Left panel: the  posterior probability distribution function $P(f_{\chi}^{\max}|\text{EM})$ of the DM mass fraction limit is derived via the linear correlation between $f_{\mathrm{\chi}}^{\max}$ and $M_{\mathrm{NS}}^{\mathrm{max}}$ in combination with $P(M_{\text{NS}}^{\max} \mid \text{EM})$.}\label{fig:5}
\end{figure}

The derived upper limits on the DM content, as shown in Fig.~\ref{fig:4} and Table~\ref{table:1}, exhibit a dependence on the chosen NM EoS models. This model dependence motivates an investigation into possible correlations between the maximum DM mass fraction in DANSs, $f^{\rm{max}}_{\chi}$, and global NS properties, with the aim of mitigating model uncertainties in the results. The search for such a correlation is focused on DANSs with a typical NS mass of $1.4~M{\odot}$. Based on the  \textit{TycM} constraints of the DM content, a significant linear correlation is found between the maximum DM mass fraction in DANSs, 
$f^{*\rm{max}}_{\chi}$, and the maximum mass of pure NSs, $M_{\mathrm{NS}}^{\mathrm{max}}$, with a Pearson correlation coefficient of $r_{1}=0.97$. The correlation between $f^{*\rm{max}}_{\chi}$ and $M_{\mathrm{NS}}^{\mathrm{max}}$ is given by
\begin{align}
f_{\chi}^{*\text{max}} = 0.1 M_{\mathrm{NS}}^{\mathrm{max}} + 0.02.
\end{align}
Applying the same correlation analysis to the DM content subject to the \textit{MaxM} constraints reveals a similarly evident, and even stronger, linear correlation, with a coefficient of $r_{2}=0.98$. The corresponding relation between $f^{\rm{max}}_{\chi}$ and $M_{\mathrm{NS}}^{\mathrm{max}}$ is
\begin{align}
f_{\chi}^{\text{max}} = 0.22 M_{\mathrm{NS}}^{\mathrm{max}} - 0.44.
\end{align}
These linear correlations are presented in the main panel of Fig.~\ref{fig:5}, alongside their confidence bands. The confidence bands are constructed using a $t$ distribution with $n-2$ degrees of freedom \cite{wooldridge2016SWCL}, which is standard for small samples as it accounts for the uncertainty in estimating the residual variance. For a linear fit $\hat{y}=ax+b$, the required variance of $\hat{y}(x)$ is given by the well-known $\Delta$ method \cite{Donaldson01021987} and takes the form
\begin{align}
\operatorname{Var}(a x+b)=\frac{n x^{2}-2x\sum x_{i}+\sum x_{i}^{2}}{n\sum x_{i}^{2}-\left(\sum x_{i}\right)^{2}},
\label{eq:var}
\end{align}
where $n$  is the sample size, and $a$, $b$ are the fitted model parameters.

Given \textit{MaxM} constraints on the maximum mass of pure NSs,
$M_{\mathrm{NS}}^{\mathrm{max}}$ (see Fig.~\ref{fig:2}), the identified linear correlation permits a further constraint on the possible range of the maximum DM mass fraction $f_{\chi}^{\mathrm{max}}$ within DANSs, thereby offering a model-independent method. Such a method is based on a reasonable ansatz that, while the \textit{MaxM} constraints are provided by the supermassive pulsars, the precise internal composition of these supermassive pulsars remains undetermined, the maximum masses predicted for both pure NSs and DANSs should fall within the probability distribution function (PDF) $P(M_{\text{NS}}^{\max} \mid \text{EM})$ \cite{romani2022psr} provided by this \textit{MaxM} constraints. Using this method, by employing the distribution of the maximum mass of pure NSs from the \textit{MaxM} constraints as the prior PDF, $P(M_{\text{NS}}^{\max} \mid \text{EM})$, and the dentified linear correlation as the conditional PDF $P(f_{\chi}^{\max} \mid M_{\text{NS}}^{\max})$, the posterior PDF $P(f_{\chi}^{\max} \mid \text{EM})$ for the DM mass fraction $f_{\chi}^{\max}$  in DANSs is derived through the integration of the conditional and prior probability densities
\begin{align}
	P(f_{\chi}^{\max} \mid \text{EM}) 
	    &= \int \mathrm{d}M_{\text{NS}}^{\max}P(f_{\chi}^{\max} \mid M_{\text{NS}}^{\max}) \nonumber \\
	    &\quad \cdot P(M_{\text{NS}}^{\max} \mid \text{EM}) \, .
	\label{eq:P1}
\end{align}
The detailed approach is as follows, the correlation between the datasets $(x_i, y_i )$ is modeled via the function $\hat{y} = f(x)$ derived from the least-squares fitting. For a given $x$, the corresponding $y$ values are described by the conditional PDF $P(y|x)$, which is formulated using a $t$ distribution with $n - 2$ degrees of freedom
\begin{align}
	P(y \mid x) 
	&= \frac{\Gamma\left(\frac{n-1}{2}\right)}{\sqrt{(n-2)\pi}\,\Gamma\left(\frac{n-2}{2}\right)} \nonumber \\
	&\quad \cdot \left(1 + \frac{(y - \widehat{y})^{2}}{(n-2)\sigma^{2} \cdot \text{Var}}\right)\label{eq:P2},
\end{align}
where $\Gamma(x)$ is the Gamma function, $\sigma^{2}$ is the population error variance (typically unknown and unbiasedly estimated by the mean squared error $S^{2}=\frac{\sum\left(y_{i}-\widehat{y}_{i}\right)^{2}}{n-2}$), and Var is the variance of the fitted model as given by Eq.~\ref{eq:var}.

The posterior PDF $P(f_{\chi}^{\max}|\text{EM})$ is subsequently obtained by the substitution of $P(M_{\text{NS}}^{\max} \mid \text{EM})$ and $P(f_{\chi}^{\max} \mid M_{\text{NS}}^{\max})$ into Eqs.~(\ref{eq:var})--(\ref{eq:P2}). This yields the PDF for the DM mass fraction in a $1.4\,M_\odot$ DANS, with the result presented in the left panel of Fig.~\ref{fig:5}. The blue shaded region corresponds to the PDF under the  \textit{TycM} constraints, indicating a value of $f_{\chi}^{\mathrm{*max}}=0.250_ {-0.011}^{+0.016}$ (68\% confidence interval for $f_{\chi}^{\mathrm{*max}}$, with lower bound $0.239$ and upper bound $0.266$). In contrast, the PDF under the \textit{MaxM} constraints, depicted by the red shaded region, is constrained to $f_{\chi}^{\max}=0.072_{-0.025}^{+0.033}$ (68\% confidence interval for $f_{\chi}^{\mathrm{max}}$, with lower bound $0.047$ and upper bound $0.105$). A comparison shows that the  \textit{TycM} constraints produce a posterior distribution with a higher median value and a narrower confidence interval than those derived from the \textit{MaxM} constraints. 
These results quantify the permissible DM mass fraction in 
$1.4~M_{\odot}$ DANSs, establishing a prior constraint for investigating their potential observational signatures.

\begin{figure}[htbp]
	\centering
	\includegraphics[width=0.48\textwidth]{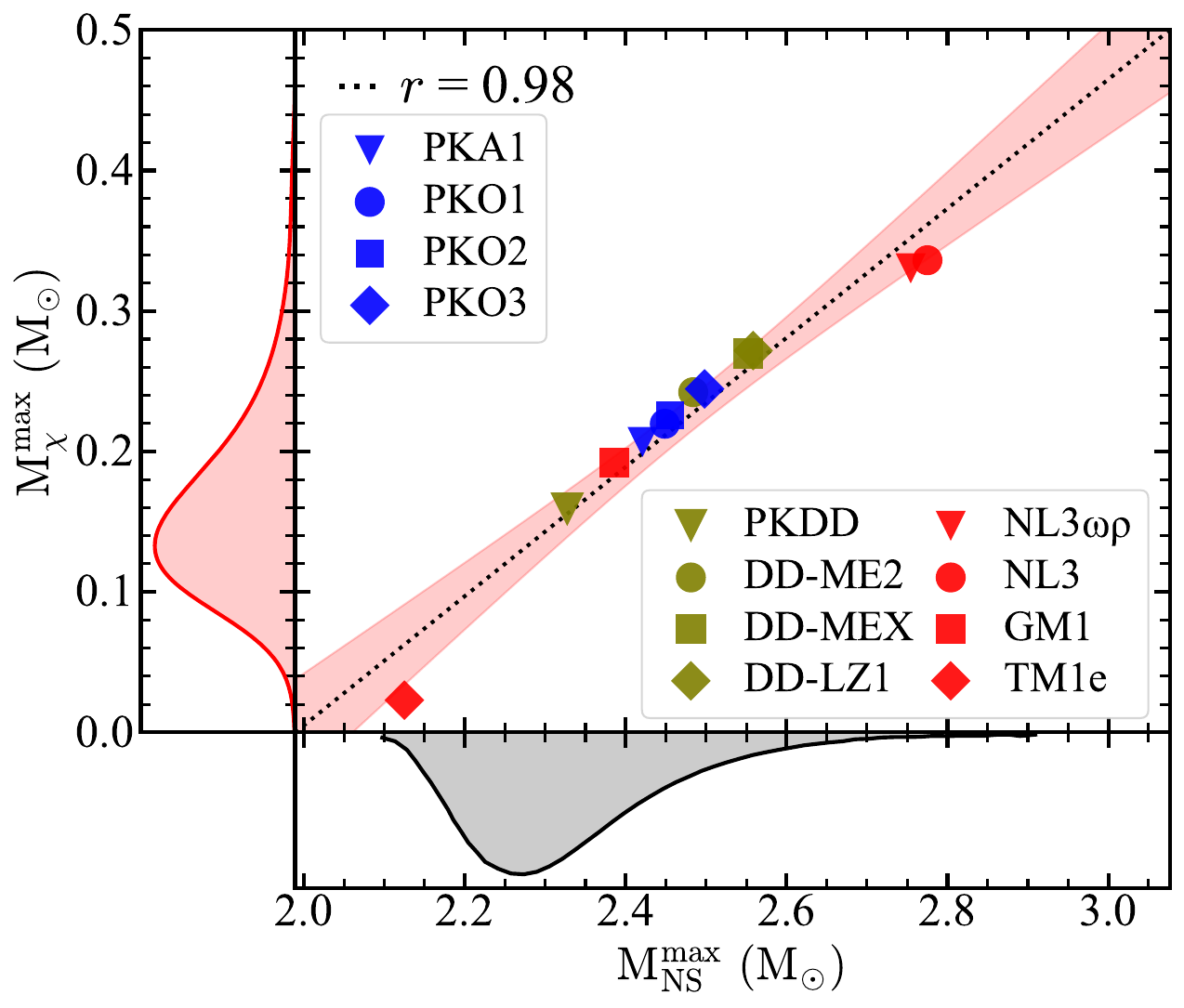}
	\caption{Linear correlation between the maximum dark matter mass $M_{\chi}^{\mathrm{max}}$ (derived from the $f_{\chi}^{\mathrm{max}}$ constrained by \textit{MaxM}) and the maximum mass $M_{\mathrm{NS}}^{\mathrm{max}}$ of a pure neutron stars. The dotted line represents the \textit{MaxM} constraints limit $M_{\chi}^{\mathrm{max}}$, with a Pearson correlation coefficient of $r = \mathrm{0.98}$. The red shaded area denotes the confidence interval.
	Bottom panel: the probability distribution function of maximum mass limit $P(M_{\text{NS}}^{\max} \mid \text{EM})$ for a pure neutron star constrained by observations \cite{romani2022psr}. 
	Left panel: the posterior probability distribution function $P(M_{\chi}^{\max}|\text{EM})$ of the maximum DM mass is derived via the linear correlation between $M_{\mathrm{\chi}}^{\max}$ and $M_{\mathrm{NS}}^{\mathrm{max}}$ in combination with $P(M_{\text{NS}}^{\max} \mid \text{EM})$.}\label{fig:6}
\end{figure}

The distribution of the DM mass fraction within a 
$1.4\,M_\odot$ DANS is analyzed in Fig.~\ref{fig:5}. Furthermore, this work aims to determine the PDF for the DM mass within a DANS at its maximum  mass, $M^{\max}_{\rm{T}}$, based on the ansatz that this mass fall within the range provided by the \textit{MaxM} constraints.  The mass-radius relation corresponding to the DM mass fraction derived from the  \textit{TycM} constraints (e.g., $f^{\mathrm{max}}_{\chi}=0.271$ in Fig.~\ref{fig:3})  fails to satisfy the \textit{MaxM} constraints. Consequently, the value of $f^{\mathrm{max}}_{\chi}$ derived under the \textit{MaxM} constraints is adopted for the subsequent analysis. Subsequently, the maximum DM mass $M^{\max}_{\chi}$ under \textit{MaxM} constraints is extracted, as shown by the solid circle on the curve for $f^{\max}_{\chi}=0.117$ in Fig.~\ref{fig:3}. The resulting values for different NM EoS models are listed in the 
$M^{\max}_{\chi}$ column of Table~\ref{table:1}. Through correlation analysis, a significant linear correlation between the maximum DM mass within DANSs, $M^{\max}_{\chi}$, and the maximum mass of a pure NS, $M^{\max}_{\mathrm{NS}}$, is also found, with a Pearson coefficient of $r = 0.98$. This correlation between the maximum DM mass within DANSs, $M^{\max}_{\chi}$, and the maximum mass of pure NSs, $M^{\max}_{\mathrm{NS}}$, which can be expressed as follows
\begin{equation}M_{\chi}^{\text{max}} = 0.46 M_{\mathrm{NS}}^{\max} - 0.91,
\label{eq:mass_correlation}
\end{equation}
and the confidence band corresponding to this correlation is presented in Fig.~\ref{fig:6}.

Based on the identified linear correlation, we proceed to analyze the probability distribution of the DM content in DANSs. The posterior PDF for the  maximum DM mass within DANSs, $P(M_{\chi}^{\mathrm{max}}|\mathrm{EM})$, is then constructed by incorporating the linear correlation between $M^{\max}_{\chi}$ and $M^{\max}_{\mathrm{NS}}$, represented by $P(M_{\chi}^{\mathrm{max}}|M_{\mathrm{NS}}^{\mathrm{max}})$, with the PDF of the \textit{MaxM} constraints $P(M_{\mathrm{NS}}^{\mathrm{max}}|\mathrm{EM})$ \cite{romani2022psr}
\begin{align}
P(M_{\chi}^{\mathrm{max}}| \mathrm{EM}) &\propto  
P(M_{\mathrm{NS}}^{\mathrm{max}} |\mathrm{EM}) \notag \\
&\quad \cdot P(M_{\chi}^{\mathrm{max}}|M_{\mathrm{NS}}^{\mathrm{max}}),
\end{align}
the explicit form of the posterior PDF $P(M_{\chi}^{\max}|\text{EM})$  is obtained by substituting  $P(M_{\text{NS}}^{\max} \mid \text{EM})$ and $P(M_{\chi}^{\max} \mid M_{\text{NS}}^{\max})$ into the framework defined by Eqs.~(\ref{eq:var})--(\ref{eq:P2}).
The resulting PDF yields a constraint on the maximum DM mass of $M_{\chi}^{\mathrm{max}}=0.150^{+0.070}_{-0.051}\ M_{\odot}$ (68\% confidence interval for $f_{\chi}^{\mathrm{max}}$, with lower bound $0.099M_{\odot}$ and upper bound $0.220M_{\odot}$), as displayed in the left panel of Fig.~\ref{fig:6}. The significance of this result can be understood as follows. Although supermassive pulsars provide the \textit{MaxM} constraints, their internal composition remains undetermined. We therefore adopt the ansatz that the predicted maximum masses of both pure NSs and DANSs are consistent with these constraints. By combining this ansatz with the identified linear correlation between $M^{\max}_{\chi}$ and $M^{\max}_{\mathrm{NS}}$, we derive the PDF    $P(M_{\chi}^{\max} \mid \mathrm{EM})$ for the DM mass within DANSs, thereby yielding a result independent of the specific NM EoS models employed.
This PDF $P(M_{\chi}^{\max} \mid \mathrm{EM})$ for the maximum DM mass provides a prior constraint for investigations into distinctive observational signatures of DANSs, such as anomalies in tidal deformability or distinctive gravitational-wave signals.

Therefore, based on the aforementioned research findings, under the  \textit{TycM} constraints, the PDF $P(f_{\chi}^{\max} \mid EM)$ for the maximum DM mass fraction $f_{\chi}^{\mathrm{max}}$ in a typical $1.4~M{\odot}$ DANS is derived, and under the \textit{MaxM} constraints, a PDF $P(M_{\chi}^{\max} \mid EM)$ for the DM component maximum mass $M_{\chi}^{\mathrm{max}}$ within DANSs is constructed. These derived distributions provide priors for the internal DM content in future studies investigating the unique observational signatures of DANSs, such as investigations into anomalies in the radius and tidal deformability of DANSs as reported in Ref.~\cite{Sun2024PhysRevD}, and studies on the distinctive frequency and duration of gravitational wave signals from DANSs as documented in Ref.~\cite{Rafiei2022PhysRevD}.

\section{Summary}\label{sec:conclusions}
This work investigates the influence of DM on the global properties of NSs, with a focus on determining the upper limit of the DM content in DANSs based on recent astrophysical observation constraints. We computed the NM EoS using twelve CDF models, and described DM with a fermionic model characterized by a particle mass $m_\chi = 1000$ MeV and a dimensionless self-interaction
strength $y=0.1$. By varying the DM mass fraction $f_{\chi}$, we analyzed its impact, finding that the predicted maximum mass $M_{\mathrm{T}}^{\mathrm{max}}$ of a DANS decreases as DM mass fraction $f_{\chi}$. Our analysis incorporates multimessenger astrophysical observations, which provide two distinct classes of constraints. The \textit{MaxM} constraints come from the supermassive pulsars mass distribution limit (with its corresponding limit from Ref.~\cite{romani2022psr}). The  \textit{TycM} constraints are provided by the NICER mass-radius measurements for PSR J0030+0451 and PSR J0437-4715, combined with data from the binary neutron star merger event GW170817. 
Under the \textit{MaxM} constraints, an upper limit on the DM content in DANSs was obtained, with the derived limit for DM mass fraction $f_{\chi}^{\max}$ shown to vary across the employed NM EoS models. For example, a value of 
$f_{\chi}^{\max}=11.7\%$ is permitted by the PKO3 model, while 
$f_{\chi}^{\max}=7.6\%$ is allowed by the PKDD model.
Similarly, under the  \textit{TycM} constraints, the theoretically attainable maximum DM content in a 
$1.4\,M_\odot$ DANS is determined and also exhibits a dependence on the NM EoS model choice.
The discrepancy in the $f_{\chi}^{\max}$ obtained from different models motivated a further investigation into possible correlations between the maximum DM mass fraction, $f_{\chi}^{\max}$, and the properties of NS.

To examine the impact of model choice on the results, we extracted the maximum DM mass fractions derived from both  \textit{TycM} and \textit{MaxM} constraints for the twelve CDF models.
By investigating possible correlations between the maximum DM mass fraction within DANSs, $f_{\chi}^{\max}$, and NS properties, a significant linear correlation was identified between this fraction, $f_{\chi}^{\max}$, and the maximum mass of a pure NS, $M_{\mathrm{NS}}^{\mathrm{max}}$. Strong correlations were found for both the  \textit{TycM} and \textit{MaxM} constrained cases, with Pearson coefficients of $r=0.97$ and $r=0.98$, respectively. Using this linear relation in conjunction with the prior PDF from the \textit{MaxM} constraints, we derived the posterior PDF for the DM mass fraction within a $1.4\,M_\odot$ DANS. 
Furthermore, this work aims to determine the PDF for the DM mass $M^{\max}_{\chi}$ within a DANS at its maximum mass, $M^{\max}_{\rm{T}}$, based on the ansatz that this mass $M^{\max}_{\rm{T}}$ fall within the range provided by the \textit{MaxM} constraints, we therefore exclusively base our investigation of the DM mass $M_{\chi}^{\max}$ in maximum mass DANSs, $M_{\mathrm{T}}^{\mathrm{max}}$, on the \textit{MaxM} constrained results. By using the values of $M_{\chi}^{\mathrm{max}}$ from Table~\ref{table:1} to examine their correlation with NS properties, we again found a significant linear correlation ($r=0.98$) between 
$M_{\chi}^{\mathrm{max}}$ and $M_{\mathrm{NS}}^{\mathrm{max}}$.
Combining this correlation with the PDF from the \textit{MaxM} constraints yields the PDF for the maximum DM mass, $P(M_{\chi}^{\mathrm{max}}| \mathrm{EM})$.
This correlation analysis leads to a constraint on the DM mass of $M_{\chi}^{\mathrm{max}}=0.150^{+0.070}_{-0.051}\ M_{\odot}$ at the 68\% confidence level. This work provides prior constraints on the DM content in DANSs that are derived from astrophysical observations and are independent of the choice of NM EoS models.

The findings of this study provide insights and a methodological framework for future investigations. Although the present analysis is based on a specific fermionic DM model with fixed parameters, the methodological framework developed here is readily extensible to broader parameter spaces and can be adopted for other DM candidate types. A probability distribution functio for the DM mass $P(M_{\chi}^{\mathrm{max}}| \mathrm{EM})$ has been constructed based on observational constraints. These results provide constraints that can be used as prior constraints for future studies that examine distinctive observational signatures, such as anomalous tidal deformabilities and distinctive gravitational-wave signals, and the global properties of DANSs.

\begin{acknowledgments}
    This work was partly supported by the National Natural Science Foundation of China (11875152), and the Fundamental Research Funds for the Central Universities, Lanzhou University (lzujbky-2023-stlt01).
    In addition, we thank Shi Yuan Ding for helpful discussions.
\end{acknowledgments}

\bibliographystyle{apsrev}
\providecommand{\noopsort}[1]{}\providecommand{\singleletter}[1]{#1}%

\end{document}